\documentclass{kluwer}    
\usepackage{psfig}

\def\eps{\varepsilon}
\def\aap{A\&A}
\def\apj{ApJ}

\def\mnras{MNRAS}
\def\aj{AJ}

\def\e{{\rm e}}
\def\me{m_\e}
\def\lesssim{\mathrel{\hbox{\rlap{\hbox{\lower4pt\hbox{$\sim$}}}\hbox{$<$}}}}
\def\gtrsim{\mathrel{\hbox{\rlap{\hbox{\lower4pt\hbox{$\sim$}}}\hbox{$>$}}}}

\def\del#1{{}}

\begin{document}                                                                                   
\begin{article}
\begin{opening}         

\title{Is the Long-Term Persistency of Circular Polarisation due to the
Constant Helicity of the Magnetic Fields in Rotating Quasar Engines?}

\runningtitle{Is the Persistency of CP due to Constant Helicity of the
Magnetic Fields?}

\author{Torsten A. \surname{En{\ss}lin}}  
\runningauthor{Torsten A. En{\ss}lin}  

\institute{Max-Planck-Institut f\"{u}r Astrophysik,
Karl-Schwarzschild-Str.1, Postfach 1317, 85741 Garching, 
Germany} 
\date{\today}

\begin{abstract}
Many compact radio sources like quasars, blazars, radio galaxies, and
micro-quasars emit circular polarisation (CP) with surprising
temporal persistent handedness. We propose that the CP is caused by
Faraday conversion (FC) of linear polarisation (LP) synchrotron light
which propagates along a line-of-sight (LOS) through helical magnetic
fields. Jet outflows from radio galaxies should have the required
magnetic helicity in the emission region due to the magnetic torque of
the accretion disc. Also advection dominated accretion flow (ADAF)
should contain magnetic fields with the same helicity. However, a jet
region seems to be the more plausible origin of CP. The proposed
scenario requires Faraday rotation (FR) to be insignificant in the
emission region.  The proposed mechanism works in electron-positron
($\e^\pm$) as well as electron-proton ($\e/{\rm p}$) plasma. In the
latter case, the emission region should consist of individual flux
tubes with independent polarities in order to suppress too strong FR
-- as it was already proposed for FR based CP generation models.  The
predominant CP is expected to mostly counter-rotate (rotation is
measured here in sky-projection) with respect to the central engine in
all cases (jet or ADAF, $\e^\pm$ or $\e/{\rm p}$ plasma) and therefore
allows to measure the sense of rotation of quasar engines.  The engine
of SgrA$^*$ is expected -- in this scenario -- to rotate clockwise and
therefore counter-Galactic, as do the young hot stars in its vicinity,
which are thought to feed SgrA$^*$ by their winds.  Generally, sources
with Stokes-$V<0$ ($V>0$) are expected to rotate clockwise
(counter-clockwise).
\end{abstract}

\end{opening}           

\section{Introduction}

CP emitting compact radio sources like quasars, blazars, radio
galaxies, and micro-quasars will be summarised in the following under
the general term quasar, assuming that a similar mechanism produces CP
in most of them. The level of CP is $\sim 1\%$ and below, usually (but
not in the case of SgrA$^*$) much lower than the level of LP. CP is
highly time variable on the one hand, but on the other hand it often
exhibits a very persistent rotational sense (per source and at nearly
all frequencies where detected), which is constant on timescales of
decades\footnote{E.g. SgrA$^*$ and GRS~1915+105.  SgrA$^*$ exhibited a
stable CP sign on a timescale (20 years) which is 5 orders of
magnitude longer than the dynamical timescale of the accretion flow
close to the black hole ($\sim$ h).  GRS~1915+105 was observed to
exhibit a LP-rotator event, but no change in the sign of CP
\cite{2002MNRAS.336...39F}. However, counter-examples seem to exist,
e.g. a CP sign reversal seems to be observed during the onset of a
radiation outburst of GRO J1655-40 in 1994
\cite{2002A&A...396..615M}.}
\cite{1984MNRAS.208..409K,1999AJ....118.1942H,2002ApJ...571..843B},
although exceptions exist. These timescales are far in excess of the
dynamical timescale of the central quasar engine from which the
emission originates.

Synchrotron radiation from relativistic electrons produces LP, but
only a very small and usually negligible amount of CP. A number of
mechanisms have been proposed to explain the observed CP, among which
FC of LP to CP seems to be the most likely process
\cite{2002MNRAS.336...39F,2002A&A...388.1106B,2002ApJ...573..485R,2002A&A...396..615M}.
FC of synchrotron emission is a two-step process, since the emitted
polarisation state has to be rotated before it can be Faraday
converted into CP. This can happen by FR, as discussed by
e.g. Ref. \cite{2002A&A...388.1106B} and \cite{2002ApJ...573..485R},
or it can be done by a systematic geometrical rotation of the magnetic
field along the LOS \cite{1982ApJ...263..595H,astro-ph/0212387}.

The motivation for the standpoint adopted here, the assumption that
the rotation of LP is a geometric effect rather than FR, is the
observational fact that CP changes are very rare, much more rare than
LP rotator events, and that a predominant CP sign seems to be
persistent in many sources.  As explained in the following, this
observational fact requires in the FR based models a constant magnetic
polarity in the emission region over timescales far in excess of the
dynamical timescale of the central engine, since the sign of CP
depends on the magnetic field polarity in such models.  In the
geometrical model discussed here, the CP sign is fully determined by
the stable rotational sense of the central engine.

\section{Magnetic helicity and CP handedness}

\begin{figure}[t]
\begin{center}
\psfig{figure=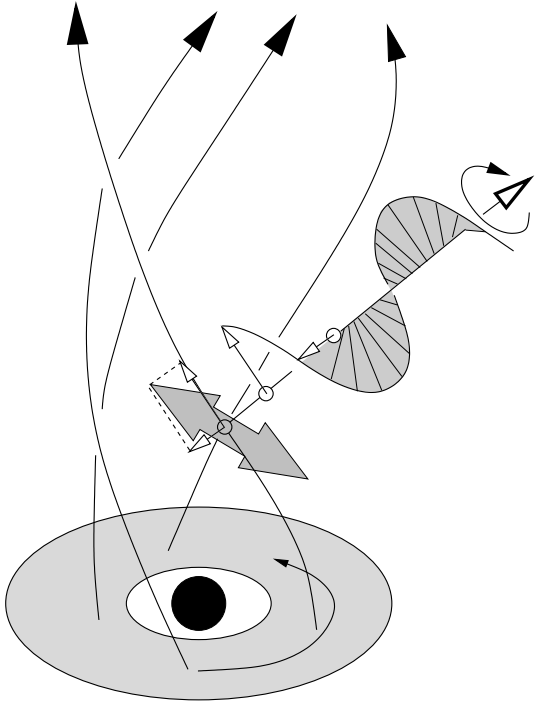,width=0.6\textwidth,angle=0}
\end{center}
\caption[]{\label{fig:jet} Sketch of a possible geometry of a jet
source. Shown is a positively rotating accretion disk around a central
black hole from which a jet is launched (not shown). Helical field
lines of the jet are drawn.  The linearly polarised radio emission of
energetic electrons within the back side magnetic field lines (the
thick arrow indicates the plane of linear polarisation) converts
partly into circular polarisation during its journey through the
foreground magnetic fields. This conversion is due to the fact that
radio emission, with linear polarisation parallel to the magnetic
field lines, propagates slower than radio emission, with perpendicular
linear polarisation. Our polarised radio emission has perpendicular
and parallel components, which oscillate in phase. Caused by the
different propagation speeds they get out of phase. The combination of
the out of phase oscillating components produces a circular
oscillation of the electric field, as sketched.}
\end{figure}

Before going into the mathematical formalism, we demonstrate how the
predominant sign of CP and its relation to the rotational sense of the
quasar engine can be understood qualitatively in the proposed
scenario. Fig. 1 illustrates the geometry.

Twisted magnetic fields are expected to be a natural ingredient of any
quasar central engine due to the rotational nature of the accretion
flow.  A helically twisted magnetic field is expected to be present in
jet outflows, especially if jets are magneto-hydrodynamically launched
\cite{1982MNRAS.199..883B}, but also otherwise, due to decreasing
rotation speed of the sideways expanding outflow. Further, if the
radio emission of a quasar would be produced in an ADAF flows
(although this is not the most natural scenario) twisted magnetic
fields with the same helicity as in the jet case can be expected. A
converging flow onto a black hole spins up due to angular momentum
conservation and should stretch magnetic fields so that they get the
same handedness as in the jet case \cite{astro-ph/0212387}.

Linearly polarised synchrotron light has to traverse foreground
magnetic fields, which are inclined with respect to the emitting
magnetic fields due to the magnetic helicity. The angle between the
linearly polarised radio emission and the local magnetic field
therefore deviates from 90$^\circ$. Since this does not correspond to
a fundamental electromagnetic mode in the plasma, the wave has to be
described as a superposition of the fundamental modes.

The two fundamental modes of electromagnetic waves travelling (nearly)
perpendicular to a magnetic field in a plasma are the linear
polarisation states with electric fields oscillating in a plane which
is parallel or perpendicular to the local magnetic field direction.
These two modes have different phase velocities since the plasma
electrons can respond more freely to an electric field oscillating
parallel to the background magnetic field than to an electric field
which tries to push the electrons in a direction perpendicular to the
magnetic field. Due to the stronger reaction of electrons on parallel
LP, their radiative back-reaction onto the propagating wave is stronger
compared to the perpendicular LP case.  Thereby the parallel LP
components is delayed with respect to the perpendicular one. The
resulting wave is elliptically polarised, therefore having a CP
component.

If the modification of the wave is a small effect, the handedness of
the CP is uniquely determined. As can be read off from the figure, but
also comes out of a more rigorous calculation, the rotation of the
electrical field of the observed wave is retrograde with respect to
the rotational sense of the accretion disc (rotation is measured in
all cases in the plane of the sky).  Numerical experiments indicate
that even in the case of strong modification the sign of CP often
follows this rule \cite{astro-ph/0212387}.

\section{Mathematical description\label{sec:math}}

The evolution of the Stokes polarisation parameters $I,Q,U,V$ along a
given LOS (here defining the $z$-axis) is governed by
\begin{equation}
\label{eq:master}
\frac{d}{dz} \left(
\begin{array}{c}
I \\ Q\\ U\\ V
\end{array}
 \right) = 
\left( 
\begin{array}{c}
\eps_I \\ \eps_Q\\ 0 \\ \eps_V
\end{array}
\right)
- \left( 
\begin{array}{cccc}
\kappa_I & \kappa_Q  & 0         & \kappa_V \\
\kappa_Q & \kappa_I  & \kappa_{\rm R}  & 0 \\
0        & -\kappa_{\rm R} & \kappa_I  & \kappa_{\rm C} \\
\kappa_V & 0         & -\kappa_{\rm C} & \kappa_I 
\end{array}
\right)  \left( 
\begin{array}{c}
I \\ Q\\ U\\ V
\end{array}
\right)\!\!
\end{equation}
\cite{2002A&A...388.1106B,2002ApJ...573..485R}, where $\eps_{I/Q/U/V}$
is the emissivity and $\kappa_{I/Q/U/V}$ the absorption coefficient of
radiation with Stokes parameter $I/Q/U/V$.  The emission and
absorption coefficients for Stokes-$V$ \footnote{CP, clockwise or
negatively rotating in sky-projection for $V>0$, counter-clockwise or
positively rotating for $V<0$.} are much smaller than the ones for
Stokes-$Q$ \footnote{LP, perpendicular ($Q>0$) or parallel ($Q<0$)
electric oscillation with respect to the sky-projected magnetic
field.}.  A special coordinate frame is adopted here, in which the
sky-projected magnetic field is parallel to the $y$-axis ($\vec{B} =
(0, B_y,B_z)$), which leads to vanishing synchrotron emission and
absorption coefficients for Stokes-$U$ \footnote{LP, $\pi/4$ inclined
to the sky-projected magnetic fields.}. Note, that in inhomogeneous
magnetic fields the local coordinate frame rotates along the LOS in
order to keep the $y$-axis aligned with the sky projected field. The
effect of this rotation on the Stokes parameter is included in
\begin{equation}
\kappa_{\rm R} = \kappa_{\rm F} + \Omega \mbox{ by the parameter }
\Omega = -2\,d\phi_B/dz \,,
\end{equation}
which describes the rate of LP rotation along the LOS due to the
rotation of the coordinate frame. $\phi_B$ is the position angle of
the sky-projected magnetic field in a fixed (non-varying) coordinate
frame.\footnote{For a purely stochastic magnetic field, $\Omega$ is a
random variable with zero mean. However, if there is a systematic
twist in the magnetic fields along the LOS $\Omega$ has a preferred
sign. The sign of $\Omega$ does not depend on the polarity of the
field, only on the handedness of the twist along the LOS.}  The FR
coefficient reads
\begin{equation}
\kappa_{\rm F} = - \frac{\tilde{\rho}_\e\, e^2\, B_z}{\pi\, \me^2\, c^4}
\lambda^2, 
\end{equation}
where $\tilde{\rho}_{\e}$ gives the effective Faraday-active leptonic
charge density.  For a non-relativistic plasma $\tilde{\rho}_{\e}$ is
identical to the leptonic charge density $\rho_{\e} = e\,(n_{\e^+} -
n_{\e^-})$, and for a relativistic plasma $\tilde{\rho}_{\e} \propto
\rho_{\e}$ with a constant of proportionality smaller than
one.\footnote{$n_{\e^\pm}$ is the $\e^\pm$ number density
respectively. In the relativistic case an identical spectral
distribution of the two species is assumed, otherwise they enter
$\tilde{\rho}_\e$ with different proportionality factors.}  The sign
of FR depends on the sign of $\rho_\e$ and on the direction of the
magnetic field component along the LOS ($B_z$).
The FC coefficient
\begin{equation}
\kappa_{\rm C} = - \frac{\tilde{n}_{\e}\,e^4\,B_y^2}{4\,\pi^2\, \me^3\,c^6}
\lambda^3 
\end{equation}
is sensitive to the total number of leptonic charge carriers
$\tilde{n}_\e \propto n_\e = n_{\e^+} + n_{\e^-}$ (with a
proportionality constant of one in the non-relativistic and smaller
than one in the relativistic case) and the square of the perpendicular
field strength $B_y^2$. Since both numbers are always positive,
$\kappa_{\rm C}$ is always negative (or zero).
Since both Faraday coefficients strongly increase with wavelength
their importance is largest at lowest frequencies, possibly explaining
why CP is often detected close to the synchrotron-self-absorption
frequency. 
In an $\e/{\rm p}$ plasma with diagonal magnetic fields ($|B_y| \sim
|B_z|$) the FR coefficient is usually larger in magnitude compared to
the FC coefficient, except for extremely high field strength or
extremely low frequencies. Both are not expected in our case.

Since there is no direct conversion of $Q$ synchrotron emission to $V$
(the corresponding matrix element in Eq. \ref{eq:master} is zero) CP
has to be produced via a two-step conversion in this setting: After
rotation (due to twist or FR) of $Q$ into $U$ it is further converted
by FC to $V$. 

In a relatively homogeneous source the sign of $V$ depends only on the
sign of $\kappa_{\rm R}$ and therefore on the polarity of the magnetic
field with respect to the LOS in the case that FR dominates, or on the
magnetic helicity, if FR is sufficiently suppressed. The total
rotational depth ($\tau_{\rm R} = \int dz\,\kappa_{\rm R}$) should be
sufficiently high on the one hand, but on the other hand it should be
not too large within the converting region, otherwise the continued
rotation changes the sign of $U$, which leads to the production of $V$
with the opposite sign and therefore cancelling of CP. To have this
condition met for FR in the right frequency range requires some level
of fine-tuning. In the case that $\kappa_{\rm R}$ is dominated by the
helical contribution, typical expected geometries give $\kappa_{\rm R}
\sim 1$, which is ideal for CP production: An effective CP production
can be achieved if both $\tau_{\rm R}$ and $\tau_{\rm C} = \int
dz\,\kappa_{\rm C}$ are simultaneously of the order one, since $V
\approx \frac{1}{6}\, \tau_{\rm F} \,\tau_{\rm C}\, Q$ for $\tau_{\rm
F} \ll 1$ and $\tau_{\rm C} \ll 1$ within an optically thin
region\footnote{For an optically thick source, using only the region
up to $\tau =1$ is a coarse approximation, leading to $V \approx
\frac{1}{6}\, \tau_{\rm F} \,\tau_{\rm C}\, Q/\tau^2$.}, and much less
than this if $\tau_{\rm F} \ge 1$ and/or $\tau_{\rm C} \ge 1$ due to
over-rotation and/or over-conversion.

{\bf FR based CP generation:} Ref. \cite{2002A&A...388.1106B} and
\cite{2002ApJ...573..485R} demonstrated that a significant CP
production can be achieved even in the case of a very large
microscopic FR coefficient if the magnetic field is mostly
stochastic. The contributions of $\kappa_{\rm F}$ with differing signs
due to magnetic field reversals can cancel each other, leaving only a
small total Faraday depth $\tau_{\rm F}$ due to an assumed weak mean
field.  The strength and sign of this mean field determines the
strength and sign of the resulting CP respectively. Although some
fine-tuning is also necessary in this scenario in order to
simultaneously have the right order of FR and FC, the larger number of
free parameters (field strength and coherence length, ratio of mean to
stochastic field components, $\tilde{\rho}_\e$, $\tilde{n}_\e$)
provides a sufficiently large parameter space to make this scenario a
plausible explanation for the observed CP for a wide range of
objects. Ref. \cite{2002ApJ...573..485R} showed that this approach can
produce CP in the case of an $\e/{\rm p}$ and also in the case of a
(charge-asymmetric) $\e^\pm$ plasma.

However, in order to explain the long-term stability of the observed
sign of $V$ it has to be assumed in their model that the weak magnetic
mean flux has to retain its polarity. This might surprise, since the
dynamical time scales in several of the observed systems are orders of
magnitudes smaller than the time-scale over which stability of the
sign of $V$ could be established.  Furthermore, the large
fluctuations in $V$ indicate strong temporal variations in the
strength of the mean field. Since the latter should be dynamically
unimportant, its stable polarity may be best explained if it results
from an environmental large scale field, which gets dragged into the
central engine. 

{\bf Magnetic helicity based CP generation} requires FR rotation to be
sufficiently suppressed.  This could be due to a vanishing leptonic
charge density ($\rho_\e$) as expected for a $\e^\pm$ plasma, or due
to the presence of aligned, separate or nested flux tubes (with a
global large scale twist) with independently oriented (but aligned)
internal magnetic fields. As Ref. \cite{2002A&A...388.1106B} and
\cite{2002ApJ...573..485R} showed for the stochastic field case,
this can strongly reduce the Faraday depth also in the case of a large
microscopic $\kappa_{\rm F}$. In their cases, some fine-tuning was
required since FR should be somehow suppressed, but not completely,
leaving to sufficient FR from $Q$ to $U$. This fine-tuning is not
required here, since the geometrical rotation provides $Q$ to $U$
rotation even in the case of fully suppressed FR.  $\kappa_{\rm C}$ is
not affected by LOS reversals of the magnetic field direction, since
it only depends on the invariant $B_y^2$.

A typical geometry is sketched in Fig. 1 \del{\ref{fig:jet}} for an
approaching jet, which is rotating positively (in sky-projection). The
rotation sense of the sky-projected magnetic fields seen by a photon
flying from its emission point (where the synchrotron process provided
it with $Q>0$) to the observer is therefore also positive, leading to
$\Omega = -2d\phi_B/dz <0$ in this geometry. Thus, in our limit
$\kappa_{\rm F} \ll \Omega$ we obtain $V\propto \Omega\,\kappa_{\rm
C}\,Q\ge 0$ since $\kappa_{\rm C} \le 0$ always. \del{The conventional
way to describe the handedness of CP is to give the sense of rotation
of the electric field in the plane of sky. This is positive or
counter-clockwise (negative or clockwise) for $V>0$ ($V<0$).} In this
picture a positively rotating approaching jet emits negatively
(clockwise) rotating circularly polarised light.  Since the counter jets
should exhibit an oppositely twisted magnetic field, it should produce
CP of the opposite sign.  However, the approaching jet dominates the
total emission due to relativistic beaming so that its CP emission
dominates the CP of both jets. We can conclude that the rotation of
the received photons of a synchrotron emitting jet source are expected
to be retrograde to -- and therefore reveal -- the sense of rotation
of the central engine, which is the rotation of the accretion disk
and/or the black hole.

\section{Discussion and outlook}

A helical magnetic field can produce circularly polarised synchrotron
emission due to FC. If FR is sufficiently suppressed the sense of CP
rotation is expected to be opposite to the one of the central engine.
The required FR suppression can be due to a charge symmetric $\e^\pm$
plasma, in which FR is absent, or due to small-scale magnetic field
polarity changes which produce cancelling FR contributions, but do not
affect FC if the different flux tubes show alignment.  The proposed
model as the following properties:
\begin{enumerate}
\item The sign of the CP does not depend on the presence of a mean
field, but on geometrical properties of the flow pattern in the
central engine of a quasar, which are fixed by the system's angular
momentum. This may explain naturally the observed sign persistence of
the CP over periods which exceed the dynamical timescales of the
system by orders of magnitude.
\item The requirement that the rotation of the angle between LP and
projected magnetic field is of the order one (not much higher, since
over-rotation reduces CP, not much lower in order to have FC
operating) within the optically thin part of the emission region, is
naturally fulfilled by the geometrical properties of jet and ADAF
flows. 
\item CP is still converted LP in this scenario, so that the
variability of CP should exceed the one of LP as observed.
\item The mechanism can work in jets and in ADAFs. It is able to
produce CP in an $\e^\pm$ and in an e/p plasma.
\item The relatively large robustness of the mechanism to variations
in the underlying source properties may explain why CP appears in a
large variety of very different systems, such as micro-quasars, low
luminosity AGNs, and powerful quasars.
\end{enumerate}
There are conditions, under which the here proposed relation between
the (predominant) rotational sense of CP and the rotation of the
central engine is not valid anymore.\footnote{Whenever the average FR
is stronger than the twist in the magnetic fields along the LOS, the
sign of the average magnetic field will determine the sign of CP as in
the models of Ref. \cite{2002A&A...388.1106B} and
\cite{2002ApJ...573..485R}. This may occur temporarily, if thermal
plasma gets mixed into an $\e^\pm$ jet. If the emission region
contains an e/p plasma, the occurrence of a mean magnetic field --
either as a statistical fluctuation, or due to some physical reason --
can increase the average Faraday depth and lead to a CP
reversal. External CP generation from LP due to environmental
scintillation also can mask the intrinsic CP signature of a
quasar. Also within the scenario in which CP is due to bending in
magnetic fields, CP sign reversals are possible, whenever the magnetic
twist within the emission region is not linked to the central engine
rotation. This may happen temporarily in violent shocks in a jet
outflow, and would suggest that CP sign reversals of this origin are
likely accompanying emission outbursts of the source. This may explain
the observed CP sign reversal in GRO J1655-40 during an outburst in
1994, as reported by \cite{2002A&A...396..615M}.} However, the rarity
of CP sign reversals indicate that such conditions should be
exceptional, if CP is due to the here proposed mechanism.\footnote{A
CP sign flip of SS~433 is repored in these proceedings
\cite{McCetaal2003}. It is possible, that this is caused by a magnetic
field reversal, since the jets of SS~433 are known to contain ions, so
that Faraday rotation might be the dominant CP generation
mechanism. However, it is also possible that the helical twist is
generating the CP, and the CP flip is caused geometrically: the
precession of the jet can lead to a geometry within the co-moving jet
material, in which the photon path towards the observer is oppositely
directed with respect to the jet flow \cite{McCetaal2003}. If this is
the case, both models, the Faraday rotation based and the magnetic
twist based, would have a CP sign flip. Further, if this is the case,
two CP reversal are expected within each precession period, so that
this scenario has a clear observational signature.  }

The proposed mechanism provides testable predictions: 
\begin{enumerate}
\item The sign of CP from a quasar measures the rotation direction
of the quasar's engine. The electric vector of radio emission should
rotate retrograde in the sky-plane with respect to the engine
rotation. Positive $V$ polarisations implies therefore positive $=$
counter-clockwise rotation (on the sky) of the engine. 
\item Thus, we expect SgrA$^*$, which exhibited $V<0$ during the last
20 years, to rotate clockwise. This is retrograde with respect to the
galactic rotation and the rotation of the molecular gas cloud in the
galactic centre. However, it has the same rotation sense as the
population of young hot HeI stars in its direct vicinity
\cite{2000MNRAS.317..348G}, which were proposed to feed SgrA$^*$ via
their stellar winds \cite{2001dscm.conf..291G}. Conservation of
angular momentum in this wind should lead to a retrograde (clockwise)
rotating accretion disk.
\item We expect the microquasars GRS~1915+105 and SS~433 (both exhibit
$V<0$) also to rotate clockwise, and we could give corresponding
expectations for the other sources with detected CP.
\item The predominant sign of CP should be temporally constant. If the
model is correct, quasar-like CP sources should only exhibit CP flips
under exceptional circumstances.
\item CP from the counter jet should have opposite handedness than CP from
the approaching jet. This prediction may allow a discrimination of the
presented model from FR based scenarios, since in the latter the same
predominant CP sign is expected from both jets for a dipole-like mean
field component. 
\end{enumerate}
If the scenario of CP production proposed here could be demonstrated
to be operatively in quasars, it would give us a tool to measure the
sense of rotation of the most powerful and enigmatic engines of the
universe -- the violent matter flows in the direct vicinity of black
holes -- by simply looking at the spin of the received photons.

\acknowledgements I acknowledge useful discussions with F.~Meyer,
S.~Heinz, H.~Spruit, R.P.~Fender, T.~Beckert, H.~Falcke, and
C.~Pfrommer. I am grateful for having the opportunity to contribute to
these proceedings without having participated in the workshop.





\end{article}
\end{document}